\documentclass[letterpaper,
               keeplastbox,   
               ]{jacow}
\usepackage{pdfpages,multirow,ragged2e} %
\makeatletter%
	\ifboolexpr{bool{xetex}}
	 {\renewcommand{\Gin@extensions}{.pdf,%
	                    .png,.jpg,.bmp,.pict,.tif,.psd,.mac,.sga,.tga,.gif,%
	                    .eps,.ps,%
	                    }}{}
\makeatother

\ifboolexpr{bool{xetex} or bool{luatex}} 
 {}                                      
 {\usepackage[utf8]{inputenc}}           

\usepackage[USenglish]{babel}

\ifboolexpr{bool{jacowbiblatex}}%
 {%
  \addbibresource{jacow-test.bib}
  \addbibresource{biblatex-examples.bib}
 }{}
\begin{document}

\title{Beamline Volume Relief Analysis for the PIP-II SSR2 Cryomodule at Fermilab
\thanks{This manuscript has been authored by Fermi Research Alliance, LLC under Contract No. DE-AC02-07CH11359 with the U.S. Department of Energy, Office of Science, Office of High Energy Physics.}}

\author{M. Parise\thanks{mparise@fnal.gov}, D. Passarelli, J. Bernardini, Fermi National Accelerator Laboratory, 60510 Batavia, IL ,USA \\
		}
	
\maketitle

\begin{abstract}
The beam volume of the Pre-Production Single Spoke Resonator type 2 (ppSSR2) cryomodule\cite{SSR2} for the Proton Improvement Plan-II (PIP-II)\cite{pip2} project will be protected against over-pressurization using a burst disk. This contribution focuses on the analysis of the relief of such trapped volume during a catastrophic scenario with multiple systems failures. An analytical model, able to predict the pressure in the beam volume depending of the various boundary conditions, has been developed and will be presented along with the results.
\end{abstract}

\section{INTRODUCTION}

\begin{figure}[!ht]
   \centering
   \includegraphics*[width=1\columnwidth]{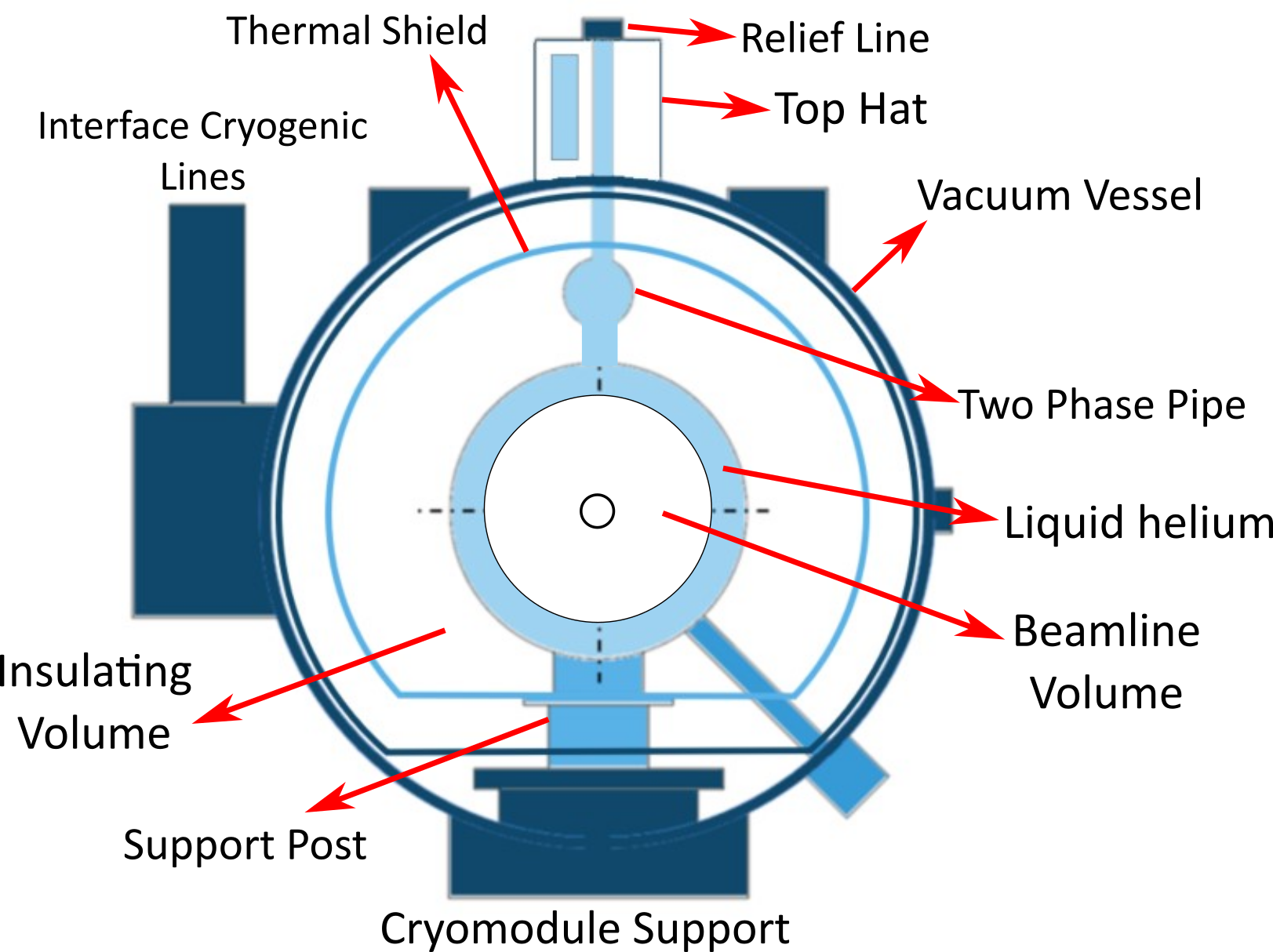}
   \caption{Schematic Section of a Generic PIP-II Cryomodule.}
   \label{fig:Scheme}
\end{figure}

Superconducting Radio Frequency (SRF) cavities are the core of linear particle accelerators like PIP-II. The new pre-production SSR2 (ppSSR2) cavities \cite{RF_Design}, \cite{Mech_design}, \cite{Fabrication} as well as Single Spoke Resonator Type 1 (SSR1) \cite{SSR1}, Low Beta (LB) \cite{LB650} and High Beta (HB) \cite{HB650} 650 MHz cavities and cryomodules \cite{PIP2CM} for PIP-II use liquid helium to cool the SRF cavities down to 2K in order to harness the superconducting advantages. Fig. \ref{fig:Scheme} shows a generic section, transversal to the beam axis, of a PIP-II cryomodule. In operation the \textit{insulating volume} and \textit{beamline volume} are under vacuum (high vacuum and ultra-high vacuum respectively). The support post is made of insulating material and interface room temperature with the beamline components at 2K. The liquid helium is contained in a circuit that encompasses the beamline components and at the operating temperature of 2K is in a two-phase state. The circuit is connected with relief devices installed outside the \textit{vacuum vessel} that are set to relieve the pressure above 4.1 bar-g when in operation. A \textit{thermal shield}, maintained at a temperature around 50K screen the beamline components from the radiation coming from the vacuum vessel. A leak between the liquid helium circuit and the \textit{beamline volume} may occur during the life of the PIP-II project. If this leak goes undetected, liquid helium may accumulate in the \textit{beamline volume}. Once the insulating vacuum is spoiled the liquid helium trapped in the \textit{beamline volume} may vaporize rapidly increasing the pressure in this space. To protect the cavities and beamline components from over-pressure, a burst disk is installed on the beamline outside of the cryomodule.

\section{INPUTS AND ASSUMPTIONS}

\begin{figure}[!ht]
   \centering
   \includegraphics*[width=1\columnwidth]{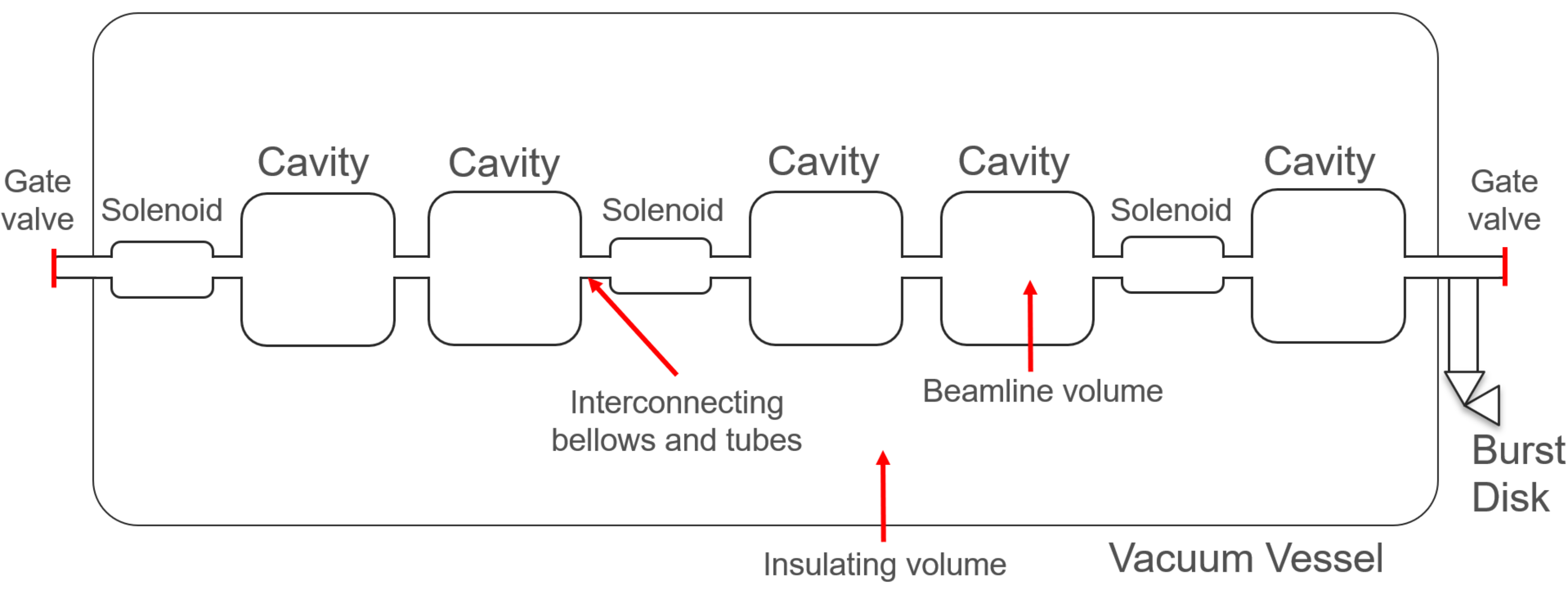}
   \caption{Schematic Geometry Used for the Analysis.}
   \label{fig:Scheme1}
\end{figure}

To purpose of this work is to analytically evaluate the pressure inside the \textit{beamline volume} as a function of the time given the applied boundary conditions under certain assumptions. Fig. \ref{fig:Scheme1} schematically shows the main elements that are recalled in this section. The inputs used for the calculation are therefore:

\begin{itemize}
    \item \underline{Geometry}: the ppSSR2 cryomodule is chosen as geometry for the analysis. It includes 5 cavities and 3 solenoids connected by bellows and tubes. The solenoid \textit{beamline volume} has the same shape and Internal Diameter (ID) of the interconnecting elements therefore it will be considered as straight pipe and the cavities are modeled as cylinders for the purpose of this analysis. The \textit{beamline volume} is considered to be closed by 2 gate valves at each ends. The \textit{beamline volume} is interfacing the \textit{insulting volume} (the helium circuit is not included. This is a conservative assumption). The complex geometry that goes from the beamline to the burst disk is decomposed in several elemental shapes for which the resistance coefficient can be easily calculated.
    \item Defect: a defect with a determined \underline{leak rate} connects the helium circuit with the textit{beamline volume}. Liquid \underline{helium} fills the volume for a \underline{certain amount of time}.
    \item Heat load to helium: the insulating vacuum is lost in \underline{1 minute of time}. As a consequence, a convective heat load is establish between the \underline{air} in that volume and the helium. The presence of a \textit{thermal shield} is also neglected and the helium therefore sees a radiative heat load from underline{300K}. The conductive heat load from the \textit{support posts} is included as well. All these heat loads are absorbed 100\% by the helium.    
\end{itemize}

All the described assumptions and the one used as part of the calculation to establish diameter changes, simplify the geometry, etc. are conservative and contribute to increase the pressure in the system. This is done on purpose so that, even if the developed mathematical model is intrinsically inaccurate because unable to capture the complexity of the geometry and heat transfer mechanisms, the result represents the worst case possible in the presence of the most unfavorable conditions.

\section{PROBLEM DESCRIPTION}

Given the underline{geometry} of the \textit{beamline volume}, the underline{leak rate} and the underline{time} during which the leak goes undetected (6 months which equals to 2 warm-ups of the cryomodule each year) is possible to calculate the volume of liquid helium at the initial condition. At t = 0 s:

\begin{itemize}
    \item the \textit{insulating volume} pressure start rising rises with an exponential law so that at t = 60 s the pressure is within 5\% of the atmospheric pressure.
    \item liquid helium and vapor coexist in the closed \textit{beamline volume} at the temperature of 2.17K and correspondent saturation pressure (all thermophysical properties are taken from the National Institute of Standards and Technology (NIST) \cite{NIST}
    \item Heat from convection, radiation and conduction starts to transfer in the closed helium system. The convection and radiation contributes depends on the temperature of the helium, the conduction contribution is orders of magnitude smaller compared to the convection therefore it is assumed constant an equal to the maximum possible throughout the calculation (8 W total)
\end{itemize}

As a result of the heat load, the temperature and pressure of the closed system increases until all liquid helium evaporate and, as the pressure reaches the burst disk set point, the system opens and the helium starts to flow outside. 

At t > 0 s, the temperature and pressure of the helium can be calculated imposing the energy balance and using the Van der Waals equations when the helium is outside the saturation vapor curve or extracting the saturation pressure at a given temperature when both the liquid and vapor phases co-exists. A code is written that solves the system of eqs. \ref{eq:1} in an iterative manner at each time step. Where: $Q_{cv}$ is the convective heat, $Q_{r}$ is the heat from radiation, $Q_{cd}$ is the heat from conduction, $M_{he}$ is the mass of helium in the system, $W$ is the helium mass flow rate exiting the system, $C_{V}$ is the specific heat constant for helium, $T_{he}$ is the helium temperature, $Q_{e}$ is the heat transported out of the system from the helium, $V$ is the helium volume, $a$ and $b$ are gas constants, $n$ is the number of moles and $R$ is the gas constant.

\begin{equation} \label{eq:1}
  \left\{\begin{array}{ll}
    (Q_{cv} + Q_r + Q_{cd})\cdot t = \\
    \,\,\,\,\,\,\,\,\,(M_{he} - W \cdot t) \cdot C_V \cdot (T_{he}(t) - T_i) + Q_{e} \cdot t \\
    (P_{he}(t)+\frac{n^2 \cdot a}{V^2} \cdot (V - n \cdot b) = n \cdot R \cdot T_{he}(t)\\
  \end{array}\right.\
\end{equation}

The quantity $W$ is calculated using eq. \ref{eq:2} from \cite{Crane} which is the Darcy formula:

\begin{equation} \label{eq:2}
W = 1.111 \cdot 10^{-6} \cdot Y \cdot d^2 \cdot \sqrt{\frac{\Delta P \cdot \rho}{K}} 
\end{equation}

\begin{figure}[!ht]
   \centering
   \includegraphics*[width=1\columnwidth]{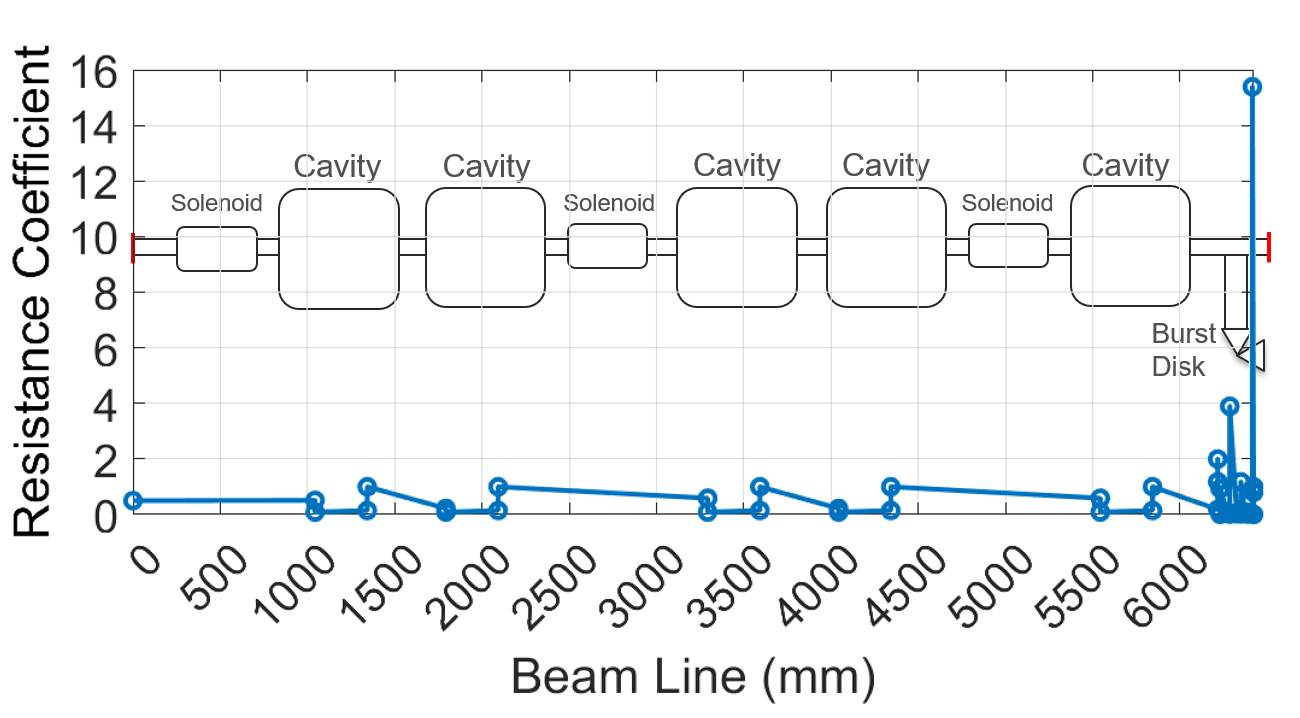}
   \caption{Schematic Geometry Used for the Analysis.}
   \label{fig:ResCoeff}
\end{figure}

Where $K$ represents the sum of the ratio between the resistance coefficients at each section and $d^4$ (where $d$ is the local outlet diameter at the end of each section) along the relief line, $\rho$ represnet the fluid density and $d$ the outlet diameter. $Y$ is the net expansion factor and in this case, for helium gas and for a choked flow, is set to 0.6 which represent a conservative assumption. Fig. \ref{fig:ResCoeff} shows the resistance coefficient along the beam line. The last portion, which includes several tees and diameter changes through valves as well as the burst disk dominate the contribution. 

\subsection{Heat Transfer}

The convective heat transfer from the air to the helium are calculated simplifying the geometry in a basic shape (cylinder) with different diameter depending on the element: cavity, solenoid or beam tube. The equation for the Nusselt number for natural convection around a cylinder is taken from \cite{bejan} and it is shown in eq. \ref{eq:3} where $Ra$ and $Pr$ are the Rayleigh an Prandtl number respectively. The thermophysical properties of air are also taken from the NIST \cite{NIST}.

\begin{equation} \label{eq:3}
Nu = \left(0.6 + \frac{0.387 \cdot Ra^\frac{1}{6}}{[1+(\frac{0.559}{Pr})^\frac{9}{16}]^\frac{8}{27}}\right)^2
\end{equation}

The contribution of the flat faces of the cylinders are also considered in the calculation using the appropriate formula for the Nusselt number (proportional to the Rayleigh and Prandtl numbers) also taken from \cite{bejan}.

The heat load to the helium is then simply calculated using eq. \ref{eq:4}:

\begin{equation} \label{eq:4}
    Q_{cv}= h \cdot \pi \cdot A \cdot \Delta T
\end{equation}

Where $h$ is the heat transfer coefficient proportional to the appropriate Nusselt number, $A$ is the surface and $\Delta T$ is the temperature difference from ambient to helium. 

The radiation heat transfer is calculated from eq. \ref{eq:5}

\begin{equation} \label{eq:5}
    Q_{r}= \frac{\epsilon_1 \cdot \epsilon_2}{\epsilon_1 + \epsilon_2 - \epsilon_1 \cdot \epsilon_2} \cdot \sigma \cdot (T_{1}^4 - T_{2}^4) \cdot A
\end{equation}

Where $\epsilon$ is the emissivity, $\sigma$ is the Boltzmann constant and $T$ is the temperature. Whilst this contribution is included in the calculation, even considering 1 layer of Multi Layer Insulation (MLI) the radiative heat transfer is negligible compared to the convective heat transfer. 

The conduction heat transfer through the solenoids is simply calculated with the appropriate material properties and the geometry (tube) using Fourier's law. As for the radiation heat transfer this contribution is included in the calculation but is negligible even compared to the radiation heat transfer.

\section{RESULTS}

\begin{figure}[ht]
   \centering
   \includegraphics*[width=0.5\textwidth]{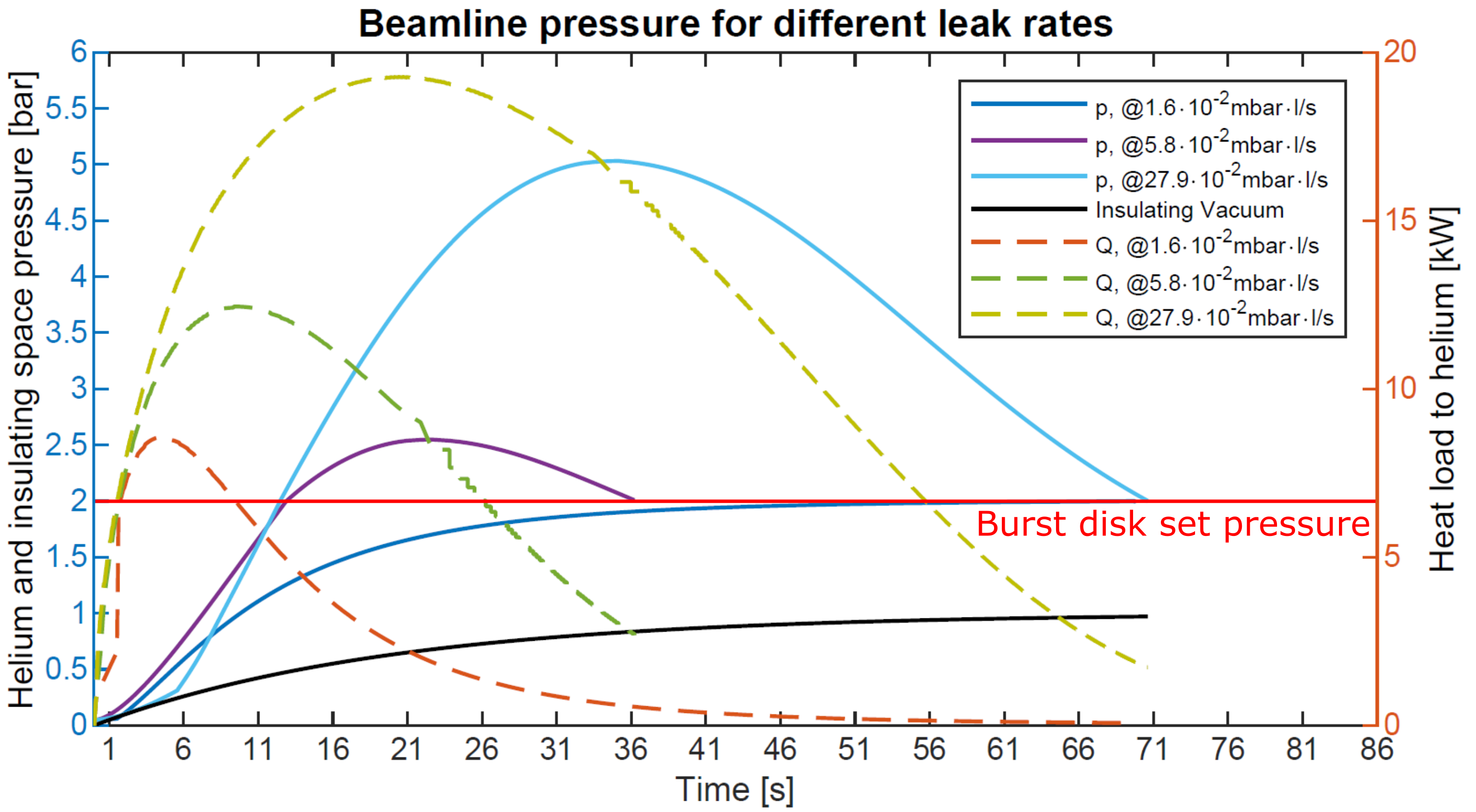}
   \caption{Pressure of and Heat Load to the Helium.}
   \label{fig:Pres}
\end{figure}

\begin{figure}[ht]
   \centering
   \includegraphics*[width=0.5\textwidth]{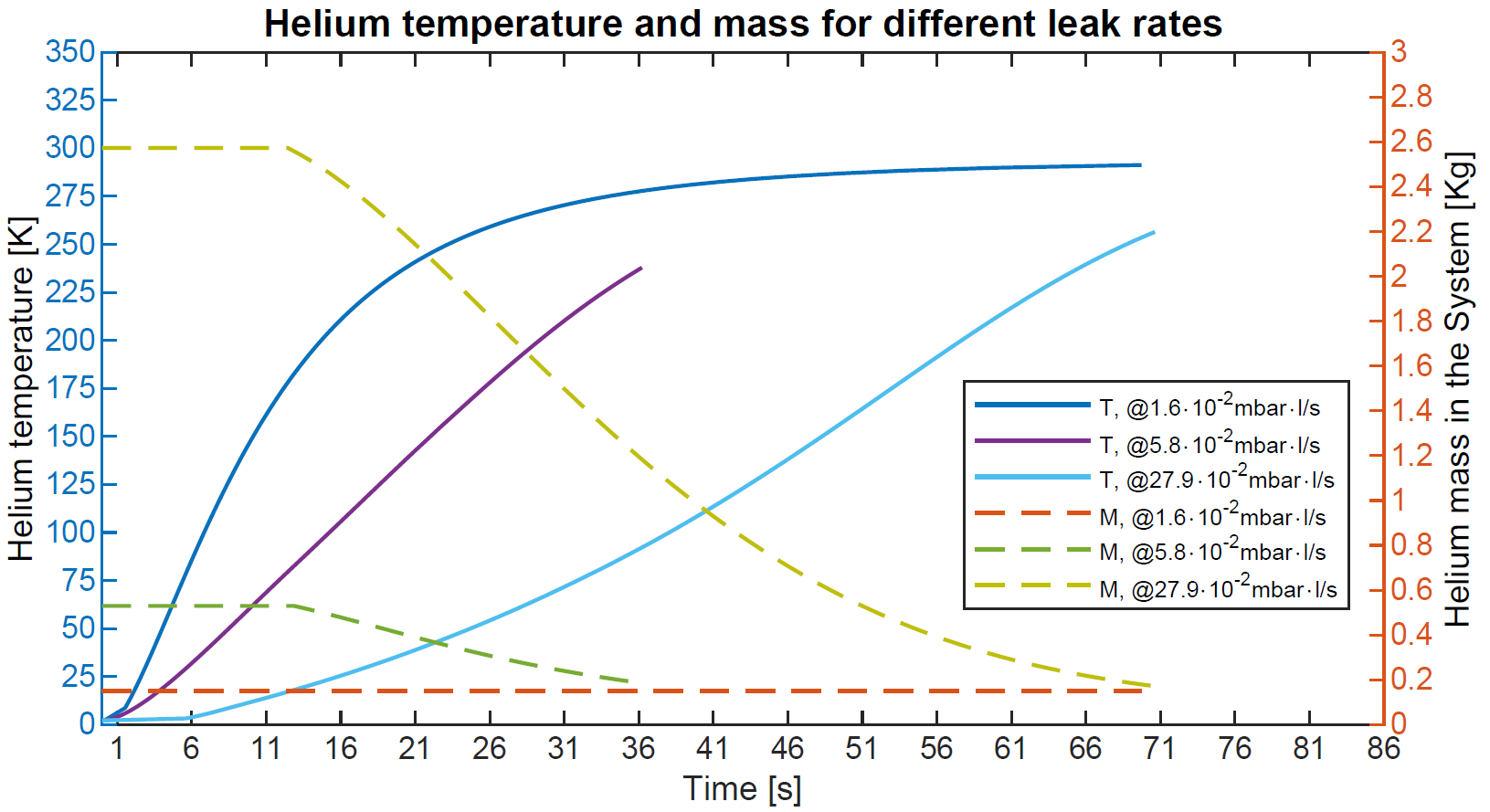}
   \caption{Temperature and Mass of Helium.}
   \label{fig:Temp}
\end{figure}

Figs. \ref{fig:Pres} and \ref{fig:Temp} shows the results of the analytical calculation. A total of 3 different leak rates are selected ranging from $1.6 \cdot 10^{-2} \frac{mbar \cdot l}{s}$ to $27.9 \cdot 10^{-2} \frac{mbar \cdot l}{s}$. The burst disk set pressure is also noted in fig. \ref{fig:Pres} (1 bar above atmosphere). The leak size below which the burst disk does not open corresponds to $1.6 \cdot 10^{-2} \frac{mbar \cdot l}{s}$. In this case the pressure rises until the burst disk set point and the temperature continue to rise until it reaches the ambient temperature. In correspondence of the biggest leak size selected, the pressure rises until approximately 5.1 bar absolute, which corresponds to 4.1 bar across the thickness of the cavity material. This represent the Maximum Allowable Working Pressure (MAWP) for the PIP-II cavities at cryogenic temperature and therefore this is the limit pressure allowed to consider the system safe. In the figures is also possible to appreciate the heat load transferred to the helium, the trend of the pressure in the insulating vacuum (Fig. \ref{fig:Pres}) and the mass of the helium in the system (Fig. \ref{fig:Temp}). 

\section{CONCLUSION}

From the results, given: the selected burst disk, the \textit{beamline volume} geometry and the conservative assumptions, the leak rate of the defect above which the safety operation of the ppSSR2 cryomodule is compromised is $27.9 \cdot 10^{-2} \frac{mbar \cdot l}{s}$. All the beamline components go through rigorous inspections at multiple steps of the fabrication and assembly operations, including leak checks performed using an helium mass spectrometer leak detector with a sensitivity better than $2 \cdot 10^{-10} \frac{mbar \cdot l}{s}$. Given that there are no dynamic loads during operation, that the applied thermal cycle goes from room temperature to cryogenic operation (thus closing any possible crack or defect), the leak checks performed before operating the machine and the time considered during which the leak goes undetected (6 months), it is safe to assume that such defect, even if present will not compromise the structure and that the burst disk is correctly sized.  

%
%
\ifboolexpr{bool{jacowbiblatex}}%
	{\printbibliography}%
	{%
	

} 
%
%


\end{document}